\documentclass[12pt]{article}
\usepackage[cp1251]{inputenc}
\usepackage{amsmath,amssymb}

\usepackage[dvips]{epsfig}
\usepackage[dvips]{graphics}
\newcommand{\rhm}{\rho_{m}}
\newcommand{\prm}{p_{m}}
\newcommand{\dph}{\dot{\phi}}

\newcommand{\vect}[1]{\boldsymbol{#1}}      
\newcommand{\veps}{\vect{\varepsilon}}

\begin{document}

\begin{center}
\Large\textbf{GRAVITATION, COSMOLOGY AND SPACE-TIME TORSION}\\
\bigskip
\normalsize A.V. Minkevich\\
\medskip
\textit{Department of Theoretical Physics, Belarussian State
University, Minsk, Belarus\\} \textit{Department of Physics and
Computer Methods, Warmia and Mazury University in Olsztyn,
Olsztyn, Poland\\}
 E-mail:MinkAV@bsu.by; awm@matman.uwm.edu.pl
\end{center}

\begin{center}
\begin{minipage}{0.8\textwidth}
\textbf{Abstract.} Poincar\'e gauge theory of gravity offers
opportunities to solve some principal problems of general
relativity theory and modern cosmology. In the frame of this
theory the gravitational interaction can have the repulsion
character in the case of usual gravitating matter with positive
values of energy density and pressure satisfying energy dominance
condition. Cosmological consequences of gravitational repulsion
are considered in the case of homogeneous isotropic models in
connection with the problem of cosmological singularity and dark
energy problem of general relativity theory. Regular Big Bang
inflationary scenario with accelerating stage of cosmological
expansion at asymptotics and the principal role of space-time
torsion in this scenario are discussed.
\end{minipage}
\end{center}

\section{Introduction}

Einsteinian general relativity theory (GR) is the base of modern
theory of gravitational interaction and relativistic cosmology. GR
allows to describe different gravitating systems and cosmological
models at widely changing scales of physical parameters. At the
same time GR possesses certain principal problems, which, in
particular, appear in cosmology. At first of all, this is the
problem of cosmological singularity (PCS): the stage of
cosmological expansion according to Einstein gravitation equations
has the beginning in the time in the past (the problem of the
beginning of the Universe in the time) and in accordance with GR
the singular state with divergent energy density and singular
metrics appears at the beginning of cosmological expansion. It is
because the gravitational interaction for usual gravitating matter
with positive values of energy density and pressure satisfying
energy dominance condition in the frame of GR as well as Newton's
theory of gravity has the character of attraction, but not
repulsion always. The PCS is particular case of general problem of
gravitational singularities of GR [1]. Note that in the frame of
GR the gravitational interaction can have the repulsion character
in the case of gravitating systems with negative pressure (for
example, scalar fields in inflationary cosmology). However, the
PCS can not be solved by taking into account such systems.
According to widely known opinion, the solution of PCS has to be
connected with quantum gravitational effects beyond Planckian
conditions, when the energy density surpasses the Planckian one. A
number of particular regular cosmological solutions was obtained
in the frame of candidates to quantum gravitation theory - string
theory/M-theory and loop quantum gravity. Radical ideas connected
with notions of strings, branes, extra-dimensions, space-time foam
etc are used in these theories (some features of these solutions
are discussed in [2,3]).

As it was shown in a number of papers (see [2-4] and Refs herein)
the gauge approach in theory of gravitational interaction offers
opportunities to solve the PCS in the frame of usual
field-theoretical description of gravity in 4-dimensional physical
space-time, where constructive Einsteinian definitions of
space-time notions are valid locally. The structure of physical
space-time in the framework of gauge approach to gravitation,
generally speaking, is more complicated in comparison with GR. So,
in the frame of the Poincar\'e gauge theory of gravity (PGTG),
which is the most important gauge theory of gravitation, the
physical space-time possesses the structure of Riemann-Cartan
continuum. Gravitational field in PGTG is described by means of
interacting metric and torsion fields. The presence of space-time
torsion can change the character of gravitational interaction by
certain physical conditions in the case of usual gravitating
matter. This fact plays the important role allowing to solve some
principal problems of GR including the PCS.

The present paper is organized by the following way. In Section 2
the question "why we need the Poincar\'e gauge theory of gravity"
is discussed. Principal moments concerning the solution of PCS in
the frame of PGTG are given in Section 3, and in Section 4 recent
results about possible solution of "dark energy problem" of GR
without dark energy on the base of PGTG are briefly discussed. In
Conclusion some possible physical consequences of space-time
torsion in connection with considered problems of GR are
discussed.

\section{Local gauge invariance principle and Poincar\'e gauge theory of gravity}

As it is known, the local gauge invariance principle is the basis
of modern theory of fundamental physical interactions. The theory
of electro-week interaction, quantum chromodynamics, Grand Unified
models of particle physics were built by using this principle.
From physical point of view, the local gauge invariance principle
establishes the correspondence between certain important
conserving physical quantities, connected according to the
Noether's theorem with some symmetries groups, and fundamental
physical fields, which have as a source corresponding physical
quantities and play the role of carriers of fundamental physical
interactions. The applying of this principle to gravitational
interaction leads, generally speaking, to generalization of
Einsteinian theory of gravitation.

At first time the local gauge invariance principle was applied in
order to build the gravitation theory by Utiyama in Ref.[5] by
considering the Lorentz group as gauge group corresponding to
gravitational interaction. Utiyama introduced the Lorentz gauge
field, which has transformation properties of anholonomic Lorentz
connection. By identifying this field with anholonomic connection
of riemannian space-time, Utiyama obtained Einstein gravitation
equations of GR by this way. The work by Utiyama [5] was
criticized by many authors. At first of all, if anholonomic
Lorentz connection is considered as independent gauge field, it
can be identified with a connection of Riemann-Cartan continuum
with torsion, but not riemannian connection  [6-8]. Moreover, if a
source of gravitational field includes the energy-momentum tensor
of gravitating matter, we can not consider the Lorentz group as
gauge group corresponding to gravitational interaction. Note that
metric theories of gravitation in 4-dimensional pseudo-riemannian
space--time including GR, in the frame of which the
energy-momentum tensor is a source of gravitational field, can be
introduced in the frame of gauge approach by the localization of
4-parametric translation group [9, 10] \footnote{Because in the
frame of gauge approach the gravitational interaction is connected
with space-time transformations, the gauge treatment to
gravitation has essential differences in comparison with
Yang-Mills fields connected with internal symmetries groups. As a
result, there are different gauge treatments to gravitational
interaction not detailed in this paper.}. By localizing
4-translations and introducing gauge field as symmetric tensor
field of second order, the structure of initial flat space-time
changes, and gauge field becomes to connected with metric tensor
of physical space-time. Because the localized translation group
leads us to general coordinate transformations, from this point of
view the general covariance of GR plays the dynamical role. At the
same time the local Lorentz group (group of tetrad Lorentz
transformations) in GR and other metric theories of gravitation
does not play any dynamical role from the point of view of gauge
approach, because corresponding Noether's invariant in these
theories is identically equal to zero [11]. The other treatment to
localization of translation group was presented in [12, 13], where
gravitation field was introduced as tetrad field in 4-dimensional
space-time with absolute parallelism. This theory is not covariant
with respect to localized tetrad Lorentz transformations, and in
fact it is intermediate step to gravitation theory with
independent gauge Lorentz field. If one means that the Lorentz
group plays the dynamical role in the gauge field theory and the
Lorentz gauge field exists in the nature, in this case we obtain
with necessity the gravitation theory in the Riemann-Cartan
space-time as natural generalization of GR (see, for example,
[14-16]). Corresponding theory is known as Poincar\'e gauge theory
of gravitation.

Gravitational gauge field variables in PGTG are the tetrad
$h^i{}_\mu$ (translational gauge field) and the Lorentz connection
$A^{ik}{}_\mu$ (Lorentz gauge field); corresponding field
strengths are the torsion tensor $S^i{}_{\mu\nu}$ and the
curvature tensor $F^{ik}{}_{\mu\nu}$ defined as
\[
S^i{}_{\mu \,\nu }  = \partial _{[\nu } \,h^i{}_{\mu ]}  -
h_{k[\mu } A^{ik}{}_{\nu ]}\,,
\]
\[
F^{ik}{}_{\mu\nu }  = 2\partial _{[\mu } A^{ik}{}_{\nu ]}  +
2A^{il}{}_{[\mu } A^k{}_{|l\,|\nu ]}\,,
\]
where holonomic and anholonomic space-time coordinates are denoted
by means of greek and latin indices respectively. As sources of
gravitational field in PGTG are energy-momentum and spin tensors.
The gravitational Lagrangian of PGTG is invariant built by means
of gravitational field strengths. The simplest PGTG is the
Einstein-Cartan theory based on gravitational Lagrangian in the
form of scalar curvature of Riemann-Cartan space-time [7,8,17]. In
certain sense the Einstein-Cartan theory of gravitation is
degenerate theory [18]. Like gauge Yang-Mills fields,
gravitational Lagrangian of PGTG has to include invariants
quadratic in gravitational field strengths - curvature and torsion
tensors. The including of linear in curvature term (scalar
curvature) to gravitational Lagrangian is necessary to satisfy the
correspondence principle with GR.

We will consider the PGTG with gravitational Lagrangian ${\cal
L}_{\rm G}$ given in general form containing different invariants
quadratic in the curvature and torsion tensors
\begin{eqnarray}\label{1}
{\cal L}_{\rm G}=  f_0\,
F+F^{\alpha\beta\mu\nu}\left(f_1\:F_{\alpha\beta\mu\nu}+f_2\:
F_{\alpha\mu\beta\nu}+f_3\:F_{\mu\nu\alpha\beta}\right)+
F^{\mu\nu}\left(f_4\:F_{\mu\nu}+f_5\:
F_{\nu\mu}\right)+f_6\:F^2 \nonumber \\
+S^{\alpha\mu\nu}\left(a_1\:S_{\alpha\mu\nu}+a_2\:
S_{\nu\mu\alpha}\right)
+a_3\:S^\alpha{}_{\mu\alpha}S_\beta{}^{\mu\beta}, 
\end{eqnarray}
where $F_{\mu\nu}=F^{\alpha}{}_{\mu\alpha\nu}$, $F=F^\mu{}_\mu$,
$f_i$ ($i=1,2,\ldots,6$), $a_k$ ($k=1,2,3$) are indefinite
parameters, $f_0=(16\pi G)^{-1}$, $G$ is Newton's gravitational
constant (the light velocity in the vacuum $c=1$). Although the
gravitational Lagrangian (1) includes a number of indefinite
parameters, gravitational equations of PGTG for homogeneous
isotropic models (HIM) considering below depend weekly on the
choice of quadratic part of gravitational Lagrangian by virtue of
their high spatial symmetry.

\section{Problem of cosmological singularity and PGTG}

According to observational data concerning anisotropy of cosmic
microwave background, our Universe was sufficiently homogeneous
and isotropic beginning from initial stages of cosmological
expansion. In connection with this fact, the investigation of HIM
is of greatest interest for relativistic cosmology. In the frame
of PGTG homogeneous isotropic models are described in general case
by means of three functions of time: the scale factor of
Robertson-Walker metrics $R(t)$ and two torsion functions
$S_{1}(t)$ and $S_{2}(t)$ determining the following components of
torsion tensor (with holonomic indices) [19]:
$S^1{}_{10}=S^2{}_{20}=S^3{}_{30}=S_{1}(t)$,
$S_{123}=S_{231}=S_{312}=S_{2}(t)\frac{R^3r^2}{\sqrt{1-kr^2}}\sin{\theta}$,
where spatial spherical coordinates are used and $k=+1,0,-1$ for
closed, flat and open models respectively. The functions $S_{1}$
and $S_{2}$ have different properties with respect to
transformations of spatial inversions, namely, unlike $S_{1}$ the
function $S_{2}(t)$ has pseudoscalar character.

At first we will consider HIM with vanishing pseudoscalar torsion
function (see [19, 2-4] and references herein) filled by
gravitating matter with energy density $\rho$ and pressure $p$
(the average of spin distribution is supposed to be equal to
zero). In this case gravitational equations of PGTG lead to the
following generalized cosmological Friedmann equations (GCFE)
\footnote{The second indefinite parameter $a=2a_1+a_2+3a_3$
connected with quadratic in the torsion part of Lagrangian (1) in
gravitational equations for HIM has to be equal to zero, if one
supposes that cosmological equations do not contain high
derivatives with respect to the scale factor $R(t)$.}:
\begin{equation}
\label{2}
\displaystyle{\frac{k}{R^2}+\left\{\frac{d}{dt}\ln\left[R\sqrt{\left|1+\alpha\left(\rho-
3p\right)\right|}\right]\right\}^2 }\displaystyle{ =\frac{8\pi
G}{3}\;\frac{\rho+
\frac{\alpha}{4}\left(\rho-3p\right)^2}{1+\alpha\left(\rho-3p\right)}
\, ,}
\end{equation}

\begin{equation}
\label{3}
\displaystyle{R^{-1}\,\frac{d}{dt}\left[\frac{dR}{dt}+R\frac{d}{dt}\left(\ln\sqrt{\left|1+\alpha\left(\rho
-
3p\right)\right|}\right)\right]} 
\displaystyle{=-\frac{4\pi G}{3}\;\frac{\rho+3p-
\frac{\alpha}{2}\left(\rho-3p\right)^2}{
1+\alpha\left(\rho-3p\right)}\, .}
\end{equation}
where indefinite parameter $\displaystyle
\alpha=\frac{f}{3f_0\,^2}>0$ ($f = f_1  + \frac{{f_2 }} {2} + f_3
+ f_4  + f_5  + 3f_6$) has inverse dimension of energy density.
According to gravitational equations the torsion function $S_{1}$
is
\begin{equation}
\label{4} \S_{1}=
-\frac{1}{4}\frac{d}{dt}\ln\left|1+\alpha(\rho-3p)\right|
\end{equation}
and conservation law for gravitating matter has usual form
\begin{equation}\label{5}
    \dot \rho  + 3H\left( {\rho  + p} \right) = 0,
\end{equation}
where $H=\dot{R}/R $ is the Hubble parameter and a dot denotes the
differentiation with respect to time.

If the parameter $\alpha$ tends to zero, the torsion function (4)
vanishes and GCFE (2)-(3) coincide with Friedmann cosmological
equations of GR. The difference of (2)--(3) from Friedmann
cosmological equations of GR is connected with terms containing
the parameter $\alpha$. The value of $\alpha^{-1}$ determines the
scale of extremely high energy densities. Solutions of GCFE
coincide practically with corresponding solutions of GR, if the
energy density is small $\left|\alpha(\rho-3p)\right|\ll 1$
($p\neq\frac{1}{3}\rho$). The difference between GR and PGTG can
be significant at extremely high energy densities
$\left|\alpha(\rho- 3p)\right|\gtrsim 1$, where the dynamics of
HIM depends essentially on space-time torsion
\footnote{Ultrarelativistic matter ($p=\frac{1}{3}\rho$) and
gravitating vacuum ($p=- \rho$) with constant energy density are
two exceptional systems, because GCFE (2)--(3) are identical to
Friedmann cosmological equations of GR in these cases
independently on values of energy density.}.

The structure of GCFE (2)--(3) ensures regular behavior of
cosmological solutions. In order to demonstrate this fact in the
case of inflationary cosmological models, we will consider below
HIM filled with scalar field $\phi$ minimally coupled with
gravitation and gravitating matter with equation of state in the
form $p_m=p_m(\rho_m)$ (values of gravitating matter are denoted
by means of index "m"). Then the energy density $\rho$ and the
pressure $p$ take the form
\begin{equation}
\label{6} \rho=\frac{1}{2}\dot{\phi}^2+V+\rho_m \quad (\rho>0),
\quad p=\frac{1}{2}\dot{\phi}^2-V+p_m,
\end{equation}
where $V=V(\phi)$ is a scalar field potential. Because the energy
density $\rho$ is positive and $\alpha>0$, from equation (2) in
the case $k=+1$, $0$ follows the relation:
\begin{equation}
\label{7}
Z=1+\alpha\left(\rho-3p\right)=1+\alpha\left(4V-\dph^2+\rhm-3\prm\right)\ge
0.
\end{equation}
The condition (7) is valid not only for closed and flat models,
but also for cosmological models of open type ($k=-1$) [2]. The
domain of admissible values of scalar field $\phi$, time
derivative $\dot{\phi}$ and energy density $\rho_m$ determined by
(7) is limited in space $P$ of these variables by bound $L$
defined as
\begin{equation} \label{8}
Z=0\quad \mbox{or}\quad
\dot\phi=\pm\left(4V+\alpha^{-1}+\rhm-3\prm\right)^{\frac{1}{2}}.
\end{equation}
Unlike GR at compression stage the time derivative $\dot{\phi}$
does not diverge, and by reaching the bound $L$ the transition to
the second part of cosmological solution containing the expansion
stage takes place. From cosmological equation (2) by using the
conservation law (5) follows that in space $P$ there are extremum
surfaces, in points of which the Hubble parameter vanishes [2].
Extremum surfaces play the role of "bounce surfaces", because the
time derivative of the Hubble parameter is positive on the
greatest part of these surfaces in the case of scalar field
potentials applying in chaotic inflation [2,4]. All cosmological
solutions have bouncing character and are regular with respect to
metrics, Hubble parameter and its time derivative. If gravitating
matter satisfies standard conditions (energy density is positive,
energy dominance condition is valid), any cosmological solution is
not limited in the time, and before the expansion stage
cosmological solution contains the compression stage and regular
transition from compression to expansion. If the value of scalar
field at the beginning of cosmological expansion is sufficiently
large ($\phi \ge 1 M_p$, where $M_p$ is the Planckian mass), like
GR cosmological solution contains quasi-de-Sitter inflationary
stage and post-inflationary stage with oscillating scalar field.
Because the de-Sitter solution is exact solution of GCFE [20],
characteristics of inflationary stage in our theory (in
particular, the duration of this stage by given initial conditions
for scalar field at the beginning of expansion) are close to that
of GR. As numerical analysis of inflationary solutions of GCFE
shows [4], the duration of transition stage from compression to
expansion is several order less than duration of inflationary
stage. By taking into account that duration of inflationary stage
is extremely small [21], we can conclude that discussed regular
cosmological solutions correspond to regular Big Bang scenario or
Big Bounce. Note that if the scale of extremely high energy
densities defined by $\alpha^{-1}$ is essentially less than the
Planckian one, the behavior of cosmological solution at the end of
inflationary stage differs from that of GR (in particular, the
Hubble parameter oscillates by changing its sign) [2,4]. After
transformation of oscillating scalar fields into ultrarelativistic
particles and transition to radiation dominated stage, the further
evolution of HIM (nucleosynthesis, transition to matter dominated
stage) practically coincides with that of GR.

Regular character of all cosmological HIM describing by GCFE is
connected with gravitational repulsion effect, which takes place
in the case of usual gravitating matter with positive energy
density at extreme conditions (extremely high energy densities and
pressures), where principal role plays the space-time torsion [3].

\section{Dark energy problem of GR and PGTG}

Unlike the PCS, which is an old cosmological problem of GR, the
dark energy problem (DEP) of GR is new problem appeared together
with discovery of the acceleration of cosmological expansion at
present epoch. By using Friedmann cosmological equations of GR in
order to explain accelerating cosmological expansion, the notion
of dark energy (or quintessence) was introduced in cosmology.
According to obtained estimations, approximately 70\% of energy in
our Universe is related with some hypothetical form of gravitating
matter with negative pressure --- dark energy --- of unknown
nature. Previously a number of investigations devoted to DEP were
carried out (see review [22]). According to widely known opinion,
the dark energy is associated with cosmological term. If the
cosmological term is related to  the vacuum energy density, it is
necessary to explain, why it has the value close to critical
energy density at present epoch (see for example [23]). Note that
by including cosmological term of corresponding value to GCFE, we
can build regular cosmology with observable accelerating expansion
stage in the frame of PGTG. However, like GR, the DEP is not
solved by this way.

As it was shown in Refs. [24,25], the PGTG offers opportunities to
solve the DEP without using the notion of dark energy. It is
because the space-time torsion in PGTG can change the character of
gravitational interaction and lead to gravitational repulsion
effect not only at extreme conditions, but also at very small
energy densities. With this purpose the HIM with two torsion
functions were built and investigated in the frame of PGTG.
Cosmological equations for such HIM include the pseudoscalar
torsion function $S_2$ with its first time derivative and contain
besides $\alpha$ also two others indefinite parameters: $b = a_2 -
a_1$ with dimension of parameter $f_0$ and dimensionless parameter
$\veps$, which is function of coefficients $f_i$ at quadratic in
the curvature terms of gravitational Lagrangian. The pseudoscalar
torsion function $S_{2}$ satisfies differential equation of second
order, and according to gravitational equations the function
$S_{1}$ can be expressed as function of the Hubble parameter, the
torsion function $S_{2}$ with its first time derivative and
parameters characterizing gravitating matter. If one supposes that
$S_{2}=0$, then the equation for $S_{2}$-function vanishes and
cosmological equations and the expression for $S_{1}$-function
take previous form given in Section 3. However, there is other
solution with not vanishing function $S_{2}$. As it was shown in
Refs. [24,25], by certain restrictions on indefinite parameters
obtained cosmological equations lead to accelerating expansion
stage at asymptotics, when physical parameters characterizing
cosmological models are sufficiently small. It is because the
pseudoscalar torsion function contains at asymptotics some
constant not vanishing value, which in the case $|\veps|\ll 1$ is:
\begin{equation}\label{9}
S_2^2  = \frac{{f_0(f_0  - b)}} {{4fb}} + \frac{{\rho  - 3p}}
{{12b }}.
\end{equation}
As a result cosmological equations at asymptotics take the form of
cosmological Friedmann equations with effective cosmological
constant induced by pseudoscalar torsion function:
 \begin{equation}\label{10}
    \frac{k} {{R^2 }} + H^2  = \frac{1} {{6b }}\left[ {\rho  + \frac{{3\,\left( {f_0  - b} \right)^2}}
         {{4f}}} \right],
\end{equation}
\begin{equation}\label{11}
    \dot H + H^2  =  - \frac{1} {{12b }}\left[ {\rho  + 3p - \frac{{3\left( {f_0  - b} \right)^2 }}
        {{2f}}} \right].
\end{equation}
By using at asymptotics the equation of state for dust matter, we
obtain that cosmological equations (10-11) lead to observable
accelerating cosmological expansion, if indefinite parameters $b$
and $\alpha$ are connected by the following way $b=[1-(2{.}8
\rho_{cr}\alpha)^{1/2}]f_0$, where critical energy density is
$\rho_{cr}=6f_0 H_0^2$  ($H_0$ is the value of the Hubble
parameter at present epoch). If we suppose that the scale of
extremely high energy densities defined by $\alpha^{-1}$ is larger
than the energy density for quark-gluon matter, but less than the
Planckian one, then we obtain the corresponding estimation for
$b$, which is very close to $f_0$.

The investigation of inflationary HIM with pseudoscalar torsion
function at extreme conditions at the beginning of cosmological
expansion shows that the PGTG allows to build totally regular
inflationary Big Bang scenario [25]. Like HIM discussed in Section
3, there are extremum surfaces in space of independent variables
$\phi$, $\dot\phi$, $S_2$, $\dot{S}_2$, $\rhm$, in the points of
which the Hubble parameter vanishes $H=0$. Extremum surfaces
depend on indefinite parameters $\alpha$, $\veps$ and in the case
of open and closed models also on the scale factor $R$ (as was
noted above, the value of $b$ depends on $\alpha$ and is close to
$f_0$). Unlike HIM with vanishing pseudoscalar torsion function,
the bounce $(\dot{H}_0>0)$ takes place only in limited domain of
extremum surfaces with negligibly small values of the function
$S_{2}$. Properties of regular inflationary solutions with
pseudoscalar torsion function differ from that without
pseudoscalar torsion function and depend essentially on indefinite
parameters $\alpha$ and $\veps$.

The regular Big Bang scenario was built in the frame of PGTG by
classical description of gravitational field. If the energy
density and values of torsion functions at transition stage from
compression to expansion are less than the Planckian ones, quantum
gravitational era was absent by evolution of the Universe. If the
Planckian conditions were realized at the beginning of
cosmological expansion, quantum gravitational corrections have to
be taken into account; however, classical cosmological equations
of PGTG ensure the regular character of the Universe evolution.

\section{Conclusion}

As follows from our consideration, the PGTG leads to certain
principal differences in comparison with GR concerning the
character of gravitational interaction for usual gravitating
matter that offers opportunities to solve some principal problems
of GR. Although the direct interaction of the torsion with
minimally coupled spinless matter is absent, corresponding
physical consequences of PGTG are connected essentially with
space-time torsion by virtue of interaction between metric and
torsion fields. According to PGTG, the domain of applicability of
GR is limited, namely in the case of cosmological HIM the domain
of admissible energy densities has upper limit determined by
$\alpha^{-1}$ and lower limit equal to $\frac{3\,\left( f_0  - b
\right)^2}{4f}$. The following question appears: by what way
obtained physical consequences of PGTG can be verified? As was
noted above, the behavior of regular inflationary cosmological
models at the end of inflationary stage, generally speaking,
differs from that of GR and depends on indefinite parameter
$\alpha$, and in the case of HIM with pseudoscalar torsion
function also on parameter $\veps$. It can be possible cause of
differences of perturbations of scalar fields at the end of
inflationary stage in comparison with GR, that has direct physical
interest in connection with observable anomalies in anisotropy of
cosmic microwave background [26]. This means that the building of
perturbations theory in inflationary HIM in the frame of PGTG is
of direct physical interest and possibly can test obtained
cosmological consequences. Obtained results can be important also
for other gravitating systems in astrophysics. In particular, the
conclusion about existence of limiting (maximum) energy density
for gravitating systems can be significant for so-called
primordial black holes limiting their admissible minimum mass
\footnote{Note that the vacuum Schwarzschild solution for metrics
with vanishing torsion is exact solution of PGTG independently on
indefinite parameters of gravitational Lagrangian (1).}. Together
with dark energy problem, the problem of the origin of not
baryonic component of dark matter is principal problem of
relativistic cosmology and astrophysics. From our consideration of
DEP given in Section 4 follows that Newton's law of gravitational
attraction has limits of its applicability and space-time torsion
can be essential in Newtonian approximation. If the torsion can
lead to physical consequences in the frame of HIM as dark energy
in GR, possibly the space-time torsion in the case of
inhomogeneous matter distribution could be important for the
solution of dark matter problem.

From our analysis given above follows that the PGTG can have the
principal meaning for theory of gravitational interaction. Note
that supergravity theory built in connection with the problem of
unification of fundamental physical interactions,  corresponds,
strongly speaking, to PGTG, but not metric theory of gravitation,
because the gauge group of supergravity theory includes the
Lorentz group. As it is known, the simplest supergravity theory
corresponds to the simplest PGTG -- Einstein-Cartan theory. If the
PGTG is correct gravitation theory, in this case quantum
gravitation theory must have as quasi-classical approximation the
gravitation theory in the Riemann-Cartan, but not
pseudo-riemannian space-time.



\begin{thebibliography}{99}

\bibitem{1} S. W. Hawking and G. F. R. Ellis  {\it The Large Scale Structure of Space-Time\/}
(Cambridge: Cambridge University Press) (1973).
\bibitem{2} A. V. Minkevich, {\it Gravitation\&Cosmology\/}, \textbf{12}, 11 (2006), ({\it Preprint\/} gr-qc/0506140).
\bibitem{3} A. V. Minkevich, {\it Acta Physica Polonica\/}, \textbf{B38}, 61 (2007), ({\it Preprint\/} gr-qc/0506123).
\bibitem{4} A. V. Minkevich and A. S. Garkun, {\it Class. Quantum Grav.\/},
{\textbf{23}}, 4237 (2006), ({\it Preprint\/} gr-qc/0506130).
\bibitem{5} R. Utiyama {\it Phys. Rev.\/}, {\bf 101}, 1597 (1956).
\bibitem{6} A. M. Brodskii, D. Ivanenko, H. A. Sokolik, {\it Zhurnal Eksp. Teor. Fiz.\/}, {\bf 41}, 1307
(1961);  {\it Acta Phys. Hungar.\/}, {\bf 14}, 21 (1962).
\bibitem{7} T. W. B. Kibble, {\it J. Math. Phys.\/}, {\bf 2}, 212
(1961).
\bibitem{8} D. W. Sciama,  In: {\it Recent Developments in GR\/} (Warsaw-New York:
Pergamon Press and PMN) (1962), p.321.
\bibitem{9} A. V. Minkevich, {\it Vestsi Akad. Nauk BSSR\/}. Ser. fiz.-mat., No.~4,
117 (1966).
\bibitem{10} R. Utiyama, T. Fukuyama, {\it Progr. Theor. Phys.\/}, {\bf 45},
612 (1971).
\bibitem{11} A. V. Minkevich, V. I. Kudin, {\it Acta Phys. Polon.\/}, {\bf B5},
335 (1974).
\bibitem{12} K. Hayashi, T. Nakano, {\it Progr. Theor. Phys.\/}, {\bf 38},
491 (1967).
\bibitem{13} K. Hayashi, T. Shirafudji, {\it Phys. Rev.\/}, {\bf D19},
3524 (1979).
\bibitem{14} J. M. Cho, {\it Phys. Rev.\/}, {\bf D14}, 3335 (1976).
\bibitem{15} F. W. Hehl, {\it In:\/} ``Cosmology and Gravitation'' (New York: Plenum
Press)(1980 ).
\bibitem{16} K. Hayashi, T. Shirafuji, {\it Progr. Theor. Phys.\/}, {\bf 64}, 866 (1980);
{\bf 64}, 1435 (1980); {\bf 64}, 2222 (1980).
\bibitem{17} A. Trautman, {\it Nature (Phys. Sci.)\/}, {\bf 242},
7 (1973 ).
\bibitem{18} A. V. Minkevich, {\it Proc.of 5th
Intern. Conf. Boyai-Gauss-Lobachevsky: Methods of Non-Euclidian
Geometry in Modern Physics\/} (Minsk: Inst. of Physics NAN
Belarus) (2006), p. 150--157, ({\it Preprint\/} gr-qc/0612115).
\bibitem{19} A. V. Minkevich, {\it Vestsi Akad. Nauk BSSR. Ser. fiz.-mat.\/}, no.~2, 87 (1980);
{\it Phys. Lett.\/}, {\bf A80}, 232 (1980).
\bibitem{20} A. V. Minkevich, {\it Phys. Lett.\/}, {\bf A95}, 422
(1983).
\bibitem{21} A. Linde {\it Particle Physics and Inflationary Cosmology\/}
(Switzerland, Chur: Harwood) (1990).
\bibitem{22} V. Sahni and A. Starobinsky, {\it Int. J. Mod. Phys.}, {\bf D15},
2105 (2006), ({\it Preprint\/} astro-ph/0610026).
\bibitem{23}  T. Padmanabhan, {\it AIP Conf.Proc.}, {\bf 861}, 179 (2006 ), ({\it Preprint\/}
astro-ph/0603114).
\bibitem{24} A. V. Minkevich, A. S. Garkun and V. I. Kudin, {\it Proc.of 5th
Intern. Conf. Boyai-Gauss-Lobachevsky: Methods of Non-Euclidian
Geometry in Modern Physics\/} (Minsk: Inst. of Physics NAN
Belarus) ( 2006), p. 150--157 ({\it Preprint\/} gr-qc/0612116).
\bibitem{25} A. V. Minkevich, A. S. Garkun and V. I. Kudin, Regular
accelerating Universe without dark energy, to be published in
Class. Quantum Grav. (2007), ({\it Preprint\/} Arxiv: 0706.1157).
\bibitem{26} C. J. Copi et al., {\it Phys. Rev.\/}, {\bf D75}, 023507 ( 2007), ({\it Preprint\/} astro-ph/0605135).

\end{thebibliography}
\end{document}